\documentclass[slac_one]{revtex4}
\usepackage{graphicx}
\usepackage{fancyhdr}
\input pubboard/babarsym   
\RequirePackage{xspace}

\newcommand{\pvec}{{\bf p}}

\newcommand{\Sf}{\ensuremath{S_f}}
\newcommand{\Cf}{\ensuremath{C_f}}

\newcommand{\DE}{\ensuremath{\Delta E}}

\providecommand{\dt}{\deltat}

\newcommand\etal{{\it et al.}}
\newcommand{\half}{\ensuremath{\frac{1}{2}}}

\newcommand{\msp}{\ensuremath{\phantom{-}}}

\newcommand{\bfig}{\begin{figure}[htbpc!]}
\newcommand{\efig}{\end{figure}}
\newcommand\bef{\begin{figure}}
\newcommand\edf{\end{figure}}

\newcommand\beq{\begin{equation}}
\newcommand\eeq{\end{equation}}
\newcommand\bear{\begin{array}}
\newcommand\enar{\end{array}}
\newcommand\beqa{\begin{eqnarray}}
\newcommand\eeqa{\end{eqnarray}}
\newcommand\ben{\begin{enumerate}}
\newcommand\een{\end{enumerate}}

\newcommand{\UfourS}{\ensuremath{\Upsilon(4S)}}

\newcommand{\etagg}{\ensuremath{\eta_{\gaga}}}
\newcommand{\etappp}{\ensuremath{\eta_{3\pi}}}

\newcommand{\etaprg}{\ensuremath{\etapr_{\rho\gamma}}}

\newcommand{\etapeppgg}{\ensuremath{\etapr_{\eta(\gamma\gamma)\pi\pi}}}
\newcommand{\etapeppppp}{\ensuremath{\etapr_{\eta(3\pi)\pi\pi}}}

   \newcommand{\rhoz}{\ensuremath{\rho^0}}

\newcommand{\fetapKz}{\ensuremath{\etapr K^0}}
\newcommand{\fetapKs}{\ensuremath{\etapr\KS}}

\newcommand{\fetapKl}{\ensuremath{\etapr\KL}}

\newcommand{\etapKs}{\ensuremath{\Bz\ra\fetapKs}}

\newcommand{\etapKl}{\ensuremath{\Bz\ra\fetapKl}}

\newcommand{\fomegaKs}{\ensuremath{\omega\KS}}

\newcommand{\fpizKs}{\ensuremath{\piz\KS}}

\newcommand{\skspiz}{\ensuremath{S_{\fpizKs}}}

\def\stwob{\ensuremath{\sin\! 2 \beta   }\xspace}

\newcommand{\fKKKs}{\ensuremath{K^+K^-K^0_{\scriptscriptstyle S}}}
\newcommand{\fphiKspi}{\ensuremath{\phi K^0_{\scriptscriptstyle S}\pi^0}}
\newcommand{\fphiKpi}{\ensuremath{\phi K \pi}}
\newcommand{\fphiKppi}{\ensuremath{\phi K^+ \pi^-}}

\begin{document}


\title{
\boldmath{$C\!P$} Violation in Hadronic Penguins at \mbox{\slshape
    B\kern-0.1em{\smaller A}\kern-0.1em B\kern-0.1em{\smaller A\kern-0.2em R}}} 

%

\author{James F. Hirschauer (for the \babar\ Collaboration)}
\affiliation{University of Colorado, Boulder, CO 80309, USA}

\begin{abstract}
We present preliminary measurements of time-dependent $C\!P$-violation
parameters in the decays 
$B^0\rightarrow\omega K^0_{\scriptscriptstyle S}$, 
$B^0\rightarrow\eta^\prime K^0$,
$B^0\rightarrow\pi^0 K^0_{\scriptscriptstyle S}$,
$B^0\rightarrow \phi K^0_{\scriptscriptstyle S}\pi^0$, and
$B^0\rightarrow K^+K^-K^0_{\scriptscriptstyle S}$, which includes the
resonant final states $\phi K^0_{\scriptscriptstyle S}$ and
$f_0(980) K^0_{\scriptscriptstyle S}$.
The data sample corresponds to the full
\mbox{\slshape B\kern-0.1em{\smaller A}\kern-0.1em B\kern-0.1em{\smaller A\kern-0.2em R}}
dataset of $467\times10^6$
$B\kern 0.18em\overline{\kern -0.18em B}{}$
pairs produced at the PEP-II asymmetric-energy $e^+e^-$ collider
at the Stanford Linear Accelerator Center.
\end{abstract}

\maketitle

\thispagestyle{fancy}

\section{INTRODUCTION} 
Measurements of time-dependent \CP\ asymmetries in \Bz\ meson decays through
$b\rightarrow c \bar{c} s$ amplitudes have provided crucial tests of the
mechanism of \CP\ violation in the Standard Model (SM) \cite{CPVobsInB}.
These amplitudes contain the leading $b$-quark couplings, given by the
Cabibbo-Kobayashi-Maskawa \cite{CKM} (CKM) flavor mixing matrix, for
kinematically allowed transitions.  Decays to charmless final states such as
$\phi\Kz$, $\piz\Kz$, $\etapr\Kz$, and $\omega\Kz$
are CKM-suppressed $b\to \qqbar s$ ($q=u,d,s$) processes dominated by a
single loop (penguin) amplitude.  This amplitude has the same weak phase
$\beta = \
arg{(-V_{cd} V^*_{cb}/ V_{td} V^*_{tb})}$ of the CKM mixing matrix
as that measured in the $b \to c \bar{c} s$ transition, but is sensitive to
the possible presence of new heavy particles in the loop \cite{Penguin}.
Due to the different non-perturbative strong-interaction properties of the
various penguin decays, the effect of new physics is expected to be channel
dependent.

The CKM phase $\beta$ is accessible experimentally through interference
between the direct decay of the $B$ meson to a \CP\ eigenstate and \BzBzb\
mixing followed by decay to the same final state.  This interference is
observable through the time evolution of the decay.  In the present study,
we reconstruct one \Bz\ from $\FourS\ra\BzBzb$, which decays to the \CP\
eigenstate \fomegaKs, \fetapKs, \fetapKl, \fpizKs, \fphiKspi, or \fKKKs\
($B_{\CP}$).  From the remaining particles in the event we also reconstruct
the decay vertex of the other $B$ meson ($B_{\rm tag}$) and identify its
flavor.  The difference $\deltat \equiv t_{\CP} - t_{\rm tag}$ of the proper
decay times $t_{\CP}$ and $t_{\rm tag}$ is obtained from the measured
distance between the decay vertices of the $B_{\CP}$ and $B_{\rm tag}$ and
the boost ($\beta\gamma=0.56$) of the $\FourS$ system.  In the \fpizKs\
analysis we compute \dt\ and its uncertainty with a geometric fit to the
$\FourS \to \BzBzb$ system taking into account the reconstructed $\KS$
trajectory, the knowledge of the average interaction point (IP)~\cite{IP},
and the average $B$ meson lifetime.  The distribution of $\deltat$ is given
by
\begin{eqnarray}
  F(\dt) &=&
        \frac{e^{-\left|\deltat\right|/\tau}}{4\tau} { 1 \mp\Delta w \pm(1-2w)
                                                   \label{eq:FCPdef}
\left[-\eta_f \Sf\sin(\deltamd\deltat) - \Cf\cos(\deltamd\deltat)\right]},\nonumber
\end{eqnarray}
where $\eta_f$ is the \CP\ eigenvalue of final state $f$,
the upper (lower) sign denotes a decay accompanied by a \Bz (\Bzb) tag, $\tau$
is the mean $\Bz$ lifetime, $\deltamd$ is the mixing frequency, 
$w$ is the mistag rate, and $\Delta w \equiv w(\Bz)-w(\Bzb)$ is the
difference in mistag rates for \Bz\ and \Bzb\ tag-side decays.  The tagged
flavor and mistag parameters $w$ and $\Delta w$ are determined with a neural
network based algorithm~\cite{babarsin2betaprd}.

A nonzero value of the parameter \Cf\ would indicate direct \CP\ violation.
In these modes we expect $\Cf=0$ and $-\eta_f\Sf= \stwob$, assuming penguin
dominance of the $b \to s$ transition and neglecting other CKM-suppressed
amplitudes with a different weak phase.  However, these CKM-suppressed
amplitudes and the color-suppressed tree diagram introduce additional weak
phases whose contributions may not be
negligible~\cite{Gross,Gronau,BN,london}.  As a consequence, the measured
$S_f$ may differ from \stwob even within the SM.  This deviation $\Delta
S_f=S_f - \stwob$ is estimated in several theoretical approaches: QCD
factorization (QCDF)~\cite{BN,beneke}, QCDF with modeled
rescattering~\cite{Cheng}, soft collinear effective theory
~\cite{Zupan}, and SU(3) symmetry~\cite{Gross,Gronau,Jonat}. The
estimates are channel dependent.  Estimates of $\Delta S$ from QCDF are in
the ranges $(0.0,0.2)$, $(-0.03,0.03)$, and $(0.01,0.12)$ for \fomegaKs,
\fetapKz, and \fpizKs, respectively \cite{beneke, Zupan, CCS}; SU(3)
symmetry provides bounds of $(-0.05,0.09)$ for \fetapKz\ and $(-0.06,0.12)$
for \fpizKs\ \cite{Jonat}.  Predictions that use isospin symmetry to
relate several amplitudes, including the $I=\frac{3}{2}$ $B\ra K\pi$ amplitude, give
an expected value for $\skspiz$ near $1.0$ instead of $\stwob$~\cite{Spiks}.  
The modification of the \CP asymmetry due to the presence of
suppressed tree amplitudes in $\Bz\to\phi(\Kp\Km)\Kz$ is at $\cal
O$(0.01)~\cite{Beneke:2005pu,Buchalla:2005us}, while at higher $\Kp\Km$
masses a larger contribution at $\cal O$(0.1) is
possible~\cite{Cheng:2005ug}. 

In these proceedings, we summarize preliminary measurements of
time-dependent \CP parameters in the aforementioned $b\to \qqbar s$
penguin-dominated $B^0$ decays.  The \fomegaKs, \fetapKz, \fpizKs, and
\fKKKs\ results are updates of previous measurements
\cite{PreviousOmK,PreviousEtapK,PreviousPizK,Previous}, while the \fphiKspi\
results are first measurements.  Detailed descriptions of each analysis are
given in Refs.~\cite{PHIKPI},~\cite{2027}, and \cite{KKKS}.

\section{DETECTOR AND DATASET}

The data used in this analysis were collected with the \babar\ detector
at the \pep2\ asymmetric-energy \epem\ storage ring operating at the
Stanford Linear Accelerator Center.  We analyze the entire \babar\ dataset
collected at the \FourS\ resonance, corresponding to an integrated
luminosity of 426~fb$^{-1}$ and $(467 \pm 5)\times 10^6$ \BB\ pairs.  

A detailed description of the \babar\ detector can be found
elsewhere~\cite{BABARNIM}.  Charged particle (track) momenta are measured
with a 5-layer double-sided silicon vertex tracker (SVT) and a 40-layer
drift chamber (DCH) coaxial with a 1.5-T superconducting solenoidal magnet.
Neutral cluster (photon) positions and energies are measured with an
electromagnetic calorimeter, which also provides partial \KL\ reconstruction.
Charged hadrons are identified with a detector of internally reflected
Cherenkov light and specific ionization measurements (\dedx) in the tracking
detectors (DCH, SVT).  Finally, the instrumented flux return of the magnet
allows discrimination of muons from pions and additional detection of \KL\
mesons.

\section{ANALYSIS TECHNIQUE}

In the \fetapKs\ and \fKKKs\ analyses we reconstruct the \KS\ in the final
states $\pi^+\pi^-$ ($K^0_{\pi^+\pi^-}$) and $\pi^0\pi^0$ ($K^0_{\pi^0\pi^0}$); in the other analyses we use only the
$\pi^+\pi^-$ final state.   Other \B-daughter candidates are reconstructed with the following decays:
$\piz\ra\gaga$; 
$\eta\ra\gaga$ (\etagg); 
$\eta\ra\pip\pim\piz$ (\etappp); 
$\etapr\ra\etagg\pip\pim$ (\etapeppgg);
$\etapr\ra\etappp\pip\pim$ (\etapeppppp); 
$\etapr\ra\rhoz\gamma$ (\etaprg), where $\rhoz\ra\pip\pim$; 
and $\omega\ra\pip\pim\piz$.  
The five final states used for \etapKs\ are 
\etapeppgg$K^0_{\pi^+\pi^-}$,
\etaprg$K^0_{\pi^+\pi^-}$,
\etapeppppp$K^0_{\pi^+\pi^-}$,
\etapeppgg$K^0_{\pi^0\pi^0}$,
and \etaprg$K^0_{\pi^0\pi^0}$.
For the \etapKl\ channel we reconstruct the \etapr\ in two modes:
\etapeppgg\ and \etapeppppp.

After applying loose selection criteria to reduce the dominant continuum
$\epem\ra\qqbar$ ($q=u,d,s,c$) background, 
we
perform an unbinned maximum likelihood (ML) fit to the data to separate
signal from background and obtain the \CP-violation parameters for each
decay channel.  As input to the ML fit, we use two kinematic variables, an
event-shape Fisher discriminant, and, in the \fomegaKs, \fphiKspi, and \fKKKs\
analyses, resonance masses and decay angles.

In all analyses but \fpizKs\ and \fetapKl, we use, as kinematic variables,
the beam-energy-substituted mass \\
$\mes \equiv \sqrt{(\half s + \pvec_0\cdot\pvec_B)^2/E_0^2 - \pvec_B^2}$
and the energy difference $\DE \equiv E_B^*-\half E_0^*$, where
$(E_0,\pvec_0)$ and $(E_B,\pvec_B)$ are the laboratory four-momenta of the
\UfourS\ and the $B_{C\!P}$ candidate, respectively, and the asterisk
denotes the \UfourS\ rest frame.  In the \fpizKs\ analysis we use $m_B$, the
invariant mass of the reconstructed $B_{C\!P}$, and $m_{\rm miss}$, the
invariant mass of the $B_{\rm tag}$ computed from the known beam energy and
the measured $B_{CP}$ momentum with mass of $B_{CP}$ constrained to the
nominal $B$ meson mass~\cite{PDG}.  In the \fetapKl\ analysis we use only
the \DE\ variable because a mass constraint on the $B$ meson during the
vertex fit leaves \mes\ and \DE\ completely correlated.  

Further discrimination from continuum background is obtained with the
combination of four event-shape variables in a Fisher discriminant: the
angle with respect to the beam axis of the $B$ momentum, the angle with
respect to the beam axis of the $B$ thrust axis, and the zeroth and second
momentum-weighted angular moments $L_0$ and $L_2$, defined as $L_i = \sum_j
p_j\times\left|\cos\theta_j\right|^i,$ where $\theta_j$ is the angle with
respect to the $B$ thrust axis of daughter particle $j$, $p_j$ is its
momentum, and the sum excludes the daughters of the $B$ candidate.  
In the
\fetapKl\ analysis we also use the continuous output of the flavor tagging
algorithm as input to the discriminant.

The \fKKKs\ analysis is designed to account for variations of \CP
structure and interference over the Dalitz plot.  We use an isobar model
that includes the $K^+K^-$ resonances $f_0(980)$, $\phi(1020)$, $X_0(1550)$,
and $\chi_{c0}$ to extract $\beta_{\rm eff}$ and $A_{\CP}$ ($-C_f$) from the
amplitude and phase information over the Dalitz plot.  In the \fphiKpi\
analysis we measure 27 parameters that characterize the interference of $S$,
$P$, and $D$ $K\pi$ partial wave amplitudes.  We are able to measure the
single mixing-induced \CP-violation parameter $\beta_{\rm eff}$, which is
accessible only through the \fphiKspi\ \CP eigenstate in which we reconstruct just
$\sim60$ events, by constraining the other 26 parameters, including
$A_{\CP}$ for each partial wave, with $\sim800$ events from the \fphiKppi\
self-tagging final state.

\section{RESULTS}
\label{sec:results}
The preliminary fit results for signal event yields and \CP parameters are
shown in Table \ref{tab:results}.  We report separate results for \fetapKs\
and \fetapKl\ in addition to the combined \fetapKz\ results.  The \fKKKs\
results comes from the high-mass, non-resonant region of the Dalitz plot
($m_{K^+K^-}>1.1~{\rm GeV}$).  The total yield in the low-mass region of the
Dalitz plot ($m_{K^+K^-}<1.1~{\rm GeV}$), which are mostly $\phi\KS$ and
$f_0(980)\KS$ events, is $421\pm25$.  The \fphiKspi\ yield is the total for
all partial waves; each $\phi K^*_J$ yield is the sum of \fphiKspi\ and
\fphiKppi\ final states events since both contribute to the determination of each
direct \CP parameter $A_{\CP}$.

All $S_f$ and $\beta_{\rm eff}$ results are consistent with the value of
$\stwob$ measured in $b\rightarrow c \bar{c} s$ decays
\cite{babars2b,belles2b}.  The current world averages are
$\stwob=0.67\pm0.02$ and $\beta=0.37\pm0.02$.
All $C_f$ and $A_{\CP}$ results are consistent
with zero direct \CP-violation.  These \fKKKs\ results favor $\beta_{\rm
  eff}\simeq0.37$ and rule out at $4.8\sigma$ the solution
$\frac{\pi}{2}-\beta$ from the trigonometric ambiguity of $\beta$ from the
measurement of $\sin2\beta$.   All results are statistics limited.  The dominant
systematic uncertainty in the \fetapKz\ analysis is related to \CP
structure in the \BB\ background; the dominant systematic uncertainty in
the \fKKKs\ analysis is related to the Dalitz model.

\begin{table}[t]
\begin{center}
\caption{Preliminary fit results for signal yields and \CP parameters.  The
first errors are statistical and the second are systematic.  See
Sec.~\ref{sec:results} for explanation of results.}
\begin{tabular}{|l|c|c|c|}
\hline 
\textbf{Mode} & \textbf{Signal Yield} & \boldmath{$~~~~~~~~-\eta_f S_f~~~~~~~~$} & \boldmath{$~~~~~~~~C_f~~~~~~~~$}\\
\hline
\fomegaKs   & $~163\pm18$ & $0.55^{+0.26}_{-0.29}\pm0.02$ & $   -0.52^{+0.22}_{-0.20}\pm0.03$\\
\hline
\fetapKz    & $2515\pm69$ & $0.57\pm0.08\pm0.02$ & $   -0.08\pm0.06\pm0.02$\\
~~\fetapKs  & $1959\pm58$ & $0.53\pm0.08\pm0.02$ & $   -0.11\pm0.06\pm0.02$\\
~~\fetapKl  & $~556\pm38$ & $0.82\pm0.19\pm0.02$ & $\msp0.09\pm0.14\pm0.02$\\
\hline
\fpizKs   & $~556\pm32$ & $0.55\pm0.20\pm0.03$ & $\msp0.13\pm0.13\pm0.03$\\
\hline 
\hline 
\textbf{Mode} & \textbf{Signal Yield} & \boldmath{$\beta_{eff}$} & \boldmath{$A_{\CP}$}\\
\hline 
\fKKKs          & $1011\pm39$  & $0.52\pm0.08\pm0.03$ & $\msp0.05\pm0.09\pm0.04$\\
~~$\phi\KS$     & (see text)  & $0.13\pm0.13\pm0.02$ & $\msp0.14\pm0.19\pm0.02$\\
~~$f_0(980)\KS$ & (see text)  & $0.15\pm0.13\pm0.03$ & $\msp0.01\pm0.26\pm0.07$\\
\hline 
\fphiKspi            &$~58\pm3$ & $0.97^{+0.03}_{-0.52}$ & (see text)\\
$\phi(K\pi)^{*0}_0$  &$172\pm24$& (see text)             & $\msp0.20\pm0.14\pm0.06$ \\
$\phi K^*(892)^0$    &$535\pm38$& (see text)             & $\msp0.01\pm0.06\pm0.03$ \\
$\phi K_2^*(1430)^0$ &$167\pm21$& (see text)             & $   -0.08\pm0.12\pm0.04$ \\
\hline 
\end{tabular}
\label{tab:results}
\end{center}
\end{table}

\section{CONCLUSIONS}
We present preliminary updates of our measurements of mixing-induced
$C\!P$-violation parameters in several \\$b\ra\qqbar s$ penguin-dominated $B^0$
decays and the first measurement in the $B^0\ra\fphiKspi$ decay.  The
\fphiKspi\ analysis demonstrates a novel technique for extracting \CP
parameters from interfering amplitudes with relatively few signal events.
Significant changes to previous analyses include twice as much data for
\fomegaKs, $20\%$ more data for other analyses, and improved track
reconstruction.

\begin{acknowledgments}
I thank the organizers of ICHEP 08 for a well-organized and interesting
conference.  I also thank my \babar\ collaborators for their support,
especially my thesis advisors, Bill Ford and Jim Smith.

\end{acknowledgments}

\end{document}